\documentclass[12pt]{article}
\usepackage{amsmath}
\usepackage[cp1251]{inputenc}  
\usepackage[english,russian]{babel} 
\usepackage{amsfonts,amssymb}
\usepackage{psfrag}
\usepackage{graphicx}

\newcommand{\be}{\medskip\begin{equation}}
\newcommand{\ee}{\end{equation}  \vskip4pt}

\begin{document}

\centerline{\bf  \large The modified Newton attraction law  }

\centerline{\bf  \large and its connection with cosmological  $\Lambda$--term}

\vskip 15pt

\centerline{N.N.\,Fimin$^{1}$, V.M.\,Chechetkin$^{1,2}$}

\vskip 12pt

\centerline{{}$^1$\,\small{Keldysh Institute of Applied Mathematics of RAS, }}

\centerline{\small{125047, Miusskaya sq., 4, Moscow, Russia}}

\centerline{{}$^2$\,\small{Institute of Computer Aided Design of RAS, }}

\centerline{ \small{123056,  2nd Brestskaya st.,   19/18,  Moscow, Russia}}

\vskip 10pt
{We consider the possibility of generalizing the Newtonian law of gravity and the transition to
a general relativistic model for weak fields with the inclusion of a repulsive term identified as a cosmological constant.
The analysis includes that of the test particle's motion in a modified gravitational field of the Hilbert
metric and then the problem of the reverse transition from the post--Galilean case to the construction of a modified exact
point mass metric which includes the $\Lambda$--term.}

\vskip 19pt

 {\bf{1.~Introduction  }}

\vskip 14pt

The use of the McCrea--Milne model \cite{Milne} to describe astrophysical and cosmological problems leads to the conclusion
that it is possible to verify the existence and  quantitative estimate of the  $\Lambda$--term in the Einstein
equations \cite{Gurz1}--\cite{Gurz6}. With the help of this approach one can investigate the nature of Dark Energy (DE) and
Dark Matter (DM)  \cite{Gurz3}. In this case, it is possible to use
as a basis for the applied mathematical apparatus not only a generalization of Newton's theorem
on the equivalence of the gravitational field of a material point and a sphere of equal masses, but also the transition
in the general relativistic Einstein equations to the weak--field approximation;
also there is interesting the problem of  the reverse transition from the post--Galilean case to the construction of a modified exact  point
mass metric that formally includes the $\Lambda$--term.
As an essential example, we mention the article \cite{dopoln0}, in which authors
 have shown that in the case of a
positive cosmological constant $\Lambda >0$  inequality for Newtonian gravitational potential $\Phi_N/c^2 \ll 1$
leads to the existence of an upper
bound on the mass ($M_{max}$) and the distance ($R_{max}$) to be used in the Newtonian
limit.
Consequently, we cannot put the boundary condition for the potential at
infinity. In the solution of the Poisson
equation there will appear then a term reflecting this boundary condition at a finite
distance.

 Studying the dynamics of an individual  massive particle and the kinetics
of  many--particle system in a gravitational field, corresponding the above--mentioned metric with cosmological term,
are problems of great importance,
especially from the point of view of the possibility of experimental investigation of the effects
of influence of DE and DM on the usual baryonic matter.
Gurzadyan model \cite{Gurz1}--\cite{Gurz3} with cosmological term
is also essentially used
in various forms in hypotheses associated with attempts to construct the
generalizations of the theory of gravity. In this direction, one should point to the
Thomas--Whitehead  (TW) Gravity model
and its phenomenological applications to dark energy and some issues related to dark matter \cite{dopoln1}.
TW Gravity  arises when projective connections become dynamical fields
($(d + 1)$--dimensional action consisting of a pure projective
Gauss--Bonnet term constructed out of projective curvature quantities naturally produces an Einstein--Hilbert
term with cosmological constant and in $d = 4$ introduces a
new angular momentum constant, $J_0$, of cosmological scale). The next direction where it is used
a formal generalization of Newton's law, considered in the framework of the Gurzadyan model, is
spatially varying--$G$ gravity \cite{dopoln2}--\cite{dopoln3},
calculate the expansion of the universe under the assumptions that $G$ varies in
space and the radial size $r$ of the universe is very large (the MOND regime
of varying--$G$ gravity). The inferred asymptotic behavior turns out to be different than
that found by McCrea and Milne in the paper \cite{Milne} and new equations bear no resemblance to those
of the relativistic case. In this cosmology, the scale factor $R(t)$ increases linearly with
time $t$, the radial velocity is driven by inertia, and gravity is incapable of hindering the
expansion.  When
authors \cite{dopoln2} include a repulsive acceleration $a_{DE}$ due to dark energy, the resulting universal
expansion is then driven totally by this new term and the solutions for $a_{DE}\to 0$ do not
reduce to those of the  $a_{DE}\equiv 0$  case. This is a realization of a new Thom catastrophe:
the inclusion of the new term destroys the conservation of energy and the results are
not reducible to the previous case in which energy is conserved. Thus, taking into account the
repulsive term in the general representation of the gravitational potential
sometimes leads in certain cases to  paradoxical--looking conclusions,
the interpretation of which requires an unorthodox physical intuition.

In this paper, we
consider the possibility of modifying the Newtonian gravity  law, taking into account the weak  anti--attractive
interaction (Gurzadyan repulsive forces),
and then draw conclusions that can be obtained from this fact by passing to the general relativistic level (in particular,
for the two--body problem).

\vskip 19pt

 {\bf{2.~Modification of the Newtonian law of attraction and its explanation based on the transition to the post--Galilean
approximation}}
\vskip 14pt

As follows from the results of \cite{Milne}, the equation of purely radial motion of the test part is $D{v}/Dt=-GM/r^2$
on the surface of an expanding sphere of mass
$M(r,t)=(4\pi/3)r^3\rho(t)$      has the form
\begin{equation}
\frac{d^2R}{dt^2}=-\frac{4\pi G}{3} {\rho(t)R},~~~R(t)=F_0r(t),~~
\label{1}
\end{equation}
$$
\:v(t)\equiv \frac{dr}{dt}=rN(t),~~
\:N(t)=-\frac{1}{3\rho(t)}\frac{d\rho}{dt}.
$$
The most general form of the law of gravitational interaction, which satisfies equation (1), and takes into account
the fact established by Newton on the identity of the dynamics of a test particle in the external field of a sphere and the corresponding
point mass at the center of the sphere, was deduced by V. Gurzadyan in \cite{Gurz1} and
 has the following form
\begin{equation}
F(r)=Ar^{-2}+Br,~~~A,B={\rm const}_{1,2}.
\end{equation}
\noindent
 The function $F(r)$ is a general solution of the equation $({r^2/2})d^2F/dr^2+rdF/dr-F(r)=0$.
 In order to obtain the values of the constants $A,B$, one should turn to the solution of the Einstein equations
 the gravitational field for the static radial metric of a massive point particle   $R_{\beta\alpha}=\Lambda g_{\beta\alpha}$,
 $T^\alpha_\beta  \equiv 0$.
 We have a solution of the Hilbert--Schwarzschild problem (for point mass $m$) with the cosmological
 term:
 \begin{equation}
 {}_{HS}g_{00}\equiv\exp(\nu(r))=1-2GM/(c^2r)-\Lambda r^2/3,~~~
\end{equation}
$$
 {}_{HS}g_{rr}\equiv\exp(\mu(r))=\big(1-2GM/(c^2r)-\Lambda r^2/3 \big)^{-1}.
$$
 If we go to the weak field limit,  corresponding  Milne--McCrea  model, we obtain  $A=-GM$, $B=-\Lambda c^2/3$.
 Consequently, the form of the modified Newtonian potential is as follows: $\widetilde\Phi_N=-GM/r-\Lambda c^2r^2/6$
 (equation (\ref{1}) takes
 the form $Dv/Dt+GM(r)/r^2 = c^2\Lambda r/3$).

The correspondence of the Gurzadyan function $F(r)$ to general
 relativity in the weak field approximation is shown in \cite{Gurz2}  by methods of group theory. In \cite{Gurz3}, using this generalized
 law, it is shown that the cosmological constant quantitatively describes as expansion of the Universe through
 solutions of the Friedmann equations, and the dynamics of galaxy clusters in the weak field approximation of general
 relativity. Furthermore, in \cite{Gurz4} the value of the cosmological constant is derived using the dynamics of clusters of
 galaxies, i.e. in the case when dark matter in galaxies is described by the cosmological constant in the second term
 of the function $F(r)$. This approach, as shown in \cite{Gurz5}, is  also  able to eliminate the mystery of the Hubble constant
 (Hubble tension), i.e. differences in the magnitude of the Hubble constant obtained, on the one hand --- from the data  of the
 Planck satellite, on the other hand --- through the Hubble diagram with calibration of the  distances of galaxies (cf. \cite{RG}).

 An important difference
 between the function $F(r)$ and the Newtonian function is that for $F(r)$ the field inside the spherical shell is no
 longer force-free, as in the case of Newton's law, i.e. particle inside a spherical shell will feel it. This fact also
 finds observational support, namely, it is shown that the spherical galactic halo determines the properties of the disk of  spiral
 galaxies \cite{Kr}, which is excluded in the case of the Newtonian law.

  In view of such a remarkable support of this
 approach by means of the observational data regarding the dark matter in clusters of galaxies, the enlightening of the Hubble
 tension problem, we will conduct a detailed analysis of its further consequences.

 First of all, consider a simplified
 limiting case corresponding to the Poisson equation with the  $\Lambda$--term. For this, we consider  the action for gravity in the
  approximation of
weak relativism (in the Lagrangian representation) \cite{Fimin1}--\cite{Fimin3}:
\begin{equation}
S^L = \sum_{a, {\bf q}}
\int \frac{m_a}{2}{\dot{\bf x}}^2_{a} ({\bf q},t) \:dct\: -\:
\sum_{{a}, {\bf q}} \int m_{a} \Phi \big({\bf x}_{a} ({\bf q},t)\big)dct \: +
\end{equation}
\begin{equation}
+ \: \frac{2\kappa}{c^4}\int \int (\nabla \Phi)^2 d^3x dct \: +\: {\kappa} \int \int \Lambda d^3x dct
- \: \frac{2 \kappa \Lambda}{c^2} \int \int  \Phi d^3x dct.
\end{equation}
Varying by coordinates of particles,
we obtain the equation of motion in the post--Newtonian approximation, corresponding to the above
action: $m_{a} \ddot{\bf x}_{a}=- m_{a} \nabla  \Phi({\bf x}_{a})$
(it coincides in form with the equation of classical dynamics). Let's rewrite
the action in Euler representation by introducing partial distribution functions:
\begin{equation}
S^E = \sum_{a}
\frac{1}{2m_{a}}\int {\bf p}^2 f_{a} ({\bf x},{\bf p},t)d^3x d^3pdt\: -\:
\sum_{{a}} \int U({\bf x},t) f_{a}({\bf x}, {\bf p},t) d^3p d^3x dt \:+\:
\end{equation}
\begin{equation}
\: + \:\frac{2\kappa}{c^4}\int \int (\nabla  \Phi)^2 d^3x dt
+{\kappa} \int \int \Lambda d^3x dct
\: - \: \frac{2 \kappa\Lambda}{c^2} \int \int \Phi  d^3x dt.
\end{equation}
The inverse transformation to the Lagrangian representation can be done by substitutions
 $f_{a}  ({\bf x}, {\bf p},t) =
\sum_{{\bf q}}  \delta \big({\bf x}- {\bf x}_{a} ({\bf q},t) \big)  \delta \big({\bf p}- {\bf p}_{a} ({\bf q},t) \big)$.
 We vary $S^E$ with respect to $U$, and we get Poisson
equation with $\Lambda$--term:
\begin{equation}
\Delta \Phi= 4\pi G \sum_{a} m_{a}
\int  f_{a} ({\bf x},{\bf p},t)\:d^3p - c^2 \Lambda.
\end{equation}
The
classical Newtonian potential increases in the interval  $r \in ]0,\infty[$  monotonically  ($\Phi_N \in ]-\infty, 0[$),
but the function  $\Phi$  has a maximum
\begin{equation}
\widetilde\Phi_{max}=-\frac{3^{2/3}}{2}GM^{2/3}\Lambda^{1/3}c^{2/3}~~~{\mbox{при}}~~~r_{m}=\bigg(
\frac{GM}{3\Lambda c^2}
\bigg)^{1/3}.
\end{equation}\noindent
In this case, the gravitational force  $F$, acting on the test body, changes sign. That is, there is a change from
attraction to repulsion at  $r=r_{\pm}= \big(    3GM/(\Lambda c^2)    \big)^{1/3}$.

The Hilbert--Schwarzschild metric with  $\Lambda$--term, has the form
\begin{equation}
{}_{HS}ds^2= {}_{HS}g_{00}d(ct)^2 -{}_{HS}g_{rr}dr^2-r^2d\Omega^2,~~~d\Omega^2=d\theta^2+{\sin}^2\theta d\varphi^2.
\end{equation} \noindent
There are
two limiting cases of interest in the development of cosmological models: for $m=0$  we obtain the de Sitter metric
 ${}_{dS}ds^2=(1-r^2/r_\Lambda^2)d(ct)^2-(1-r^2/r_\Lambda^2)^{-1}dr^2-r^2d\Omega^2$,
 $r_\Lambda^2=3/\Lambda$,
 for $\nu=\mu=\Lambda=0$
we have the metric of a flat space--time ${}_{M}ds^2=dt^2-dr^2-r^2d\Omega^2$.
 In the interval $r\in (2GM/c^2, \sqrt{3/\Lambda})$
there exists a  possibility of reducing the Hilbert metric to the generalized Fock form of a weak field
$ds^2=(1+\widetilde\Phi^{(0)}/c^2)d(ct)^2-\sum_j(1-\widetilde\Phi_j(x_1,x_2,x_3)/c^2)d{x_j}^2$,
for which we consider its representation in the form of a small deviation from the Minkowski metric:
$g_{\alpha\beta}={}_{M}h_{\alpha\beta}+{}_{M}\delta g_{\alpha\beta}$.
Wherein an explicit  form of the diagonal (nonzero) components of this
deviation:
\begin{equation}
{}_{M}\delta g_{00}=-2\frac{GM}{r}-\frac{1}{3}\Lambda r^2,~~~{}_{M}\delta g_{jj}\big|_{j=1,2,3}=-2\frac{GM}{r}+\frac{\Lambda}{6}
(r^2+3x^2_j).
\end{equation} \noindent

In the non-relativistic approximation, only the time component of the perturbed metric is taken into account,
so that the equations of geodesic motion of a particle of unit mass are assumed to be of Newtonian form:
$d^2x^j/dt^2+c^2\Gamma_{00}^j=0$
(for the Hilbert metric $j=r$, $\Gamma_{00}^r=(1-2GMc^{-2}/r)GM/r^2$). In the above post--Galilean approximation with taking into
account  the cosmological term, the situation changes significantly, and the components of the pseudo--Fock metric
acquire the dependence on the spatial coordinates (3).

Test particle dynamics in the Hilbert metric with  the  $\Lambda$--term
in the equatorial plane $\theta=\pi/2$ is determined by three integrals of motion
\begin{equation}
\big(1-2GM/(c^2r)-\Lambda r^2/3\big)\frac{dt}{d\tau}=E,~~~~~r^2 \frac{d\varphi}{d\tau}=cL,
\label{12}
\end{equation}
\begin{equation}
E^2  \big(1-\frac{2GM}{c^2r}-\frac{\Lambda r^2}{3}\big)^{-1}-\frac{c^2L^2}{r^{2}}- \big(1-\frac{2GM}{c^2r}-
\frac{\Lambda r^2}{3}\big)^{-1}\big( \frac{dr}{d\tau} \big)^2=I_m, 
\label{13}
\end{equation}
$$
~~~I_{m=1}=1, ~~I_{m=0}=0.
$$
Eliminating the
proper time, we obtain the equations of the radial and meridian equatorial geodesics:
\begin{equation}
\big( \frac{dr}{dt}    \big)^2=c^2 \big(1-2GM/(c^2r)-\Lambda r^2/3\big)^{2}\times
\end{equation}
$$
\times
\bigg(1- \big( \frac{cL}{E}   \big)^2
\big(1-2GM/(c^2r)-\Lambda r^2/3\big)  \big(  \frac{1}{c^2L^2}   +\frac{1}{r^2}   \big)
\bigg),
$$
\begin{equation}
-\frac{d^2r}{d\varphi^2}+\frac{2}{r}\big( \frac{dr}{d\varphi}  \big)^2+r^3=\frac{3GM}{c^2}+\frac{GMc^2r^2}{L}-\frac{c^4r^5\Lambda}{3L}.
\end{equation}
Formally, the particle motion obeys the conservation
law in the form of the classical decompositions:  $$T_{kin}^{(eff)}+V_{pot}^{(eff)}=E^{(eff)},$$
\begin{equation}
T_{kin}^{(eff)}=\frac{1}{2}\big(  \frac{dr}{d\tau} \big)^2,
~~~
V_{pot}^{(eff)}=\frac{L^2}{2r^2}-\frac{GM}{r}-\frac{GML^2}{r^3}-\frac{c^2\Lambda}{6}(L^2+r^2),
~~~
E^{(eff)}=\frac{E^2-1}{2}.
\end{equation}
The introduction of the $\Lambda$--term, as can be seen, makes sense when considering the dynamics of a particle,
similar to the introduction of centrifugal potential. Therefore, it can be argued that cubic (with respect to $r$)
equation $V_{pot}^{(eff)}(r)=0$ can have
(depending on the relation of parameters $L,M,\Lambda$): 1) one real root (one turning point);
2) three real roots (one turning point and two apsidal points that define the region of finiteness of movement).

\vskip 19pt

 {\bf{3.~The orbits for the Hilbert metric with  the cosmological term}}
\vskip 14pt

For the Hilbert metric, the solution to the equation of orbital  quasi--finite motion
$dr/d\varphi=\sqrt{\Xi(r\:|\:E,L)}$,
$\Xi(r)\equiv \big( r^4(E^2 -m^2)/L^2 +2GMm^2 r^3/(L^2 c^2)-r^2+ 2GMr /c^2    \big)$
can be derived in terms of the $\wp$--Weierstrass function:
$r(\varphi)= r_{\Xi,1} +\Xi_r(r_{\Xi,1})/\big( 4\wp(\varphi; {\rm inv}_1, {\rm inv}_2)-\Xi_{rr} (r_{\Xi,1})/6\big)$,
where ${\rm inv}_{1,2}$ are invariants expressed in terms of the roots  $r_{\Xi,1,...,4}$ of the equation $\Xi(r)=0$.
If we take into account the presence of the  $\Lambda$--term in the metric coefficients, then the equation of the orbital motion
for a massive particle  (for $I_1=1$)
\begin{equation}
\frac{dr}{d\varphi}=\big(r^6\Lambda m^2 /(3L^2)+  r^4\big((E^2 -m^2)/L^2  +\Lambda/3 \big) +
\label{17}
\end{equation}
$$
+2GMm^2 r^3/(L^2 c^2)-r^2+ 2GMr /c^2\big)^{1/2}
$$
has a solution expressed in terms of hyper-elliptic integrals; its behavior can investigate quali\-ta\-ti\-ve\-ly,
starting
from the limiting cases. One of such cases is the analysis of the dynamics of a  particle, moving on the trajectory infinitesimally close to
the ``unperturbed''  trajectory for the Hilbert metric.
We divide the equations of the integrals of motion (\ref{13}) and (\ref{12}) by each other, and introduce a new dependent variable
 $\xi(\varphi)=1/r(\varphi)$:
\begin{equation}
\big(\frac{d\xi}{d\varphi}\big)^2=a_1\xi^3-\xi^2+  a_3,
~~~a_1\equiv \frac{2GM}{c^2},~~a_3 \equiv  \frac{E^2}{c^2L^2}+\frac{\Lambda}{3}.
\label{18}
\end{equation}
The solution to this equation is expressed
 in terms of the elliptic integral of the 1st kind $F(\xi_0 | k)$:
\begin{equation}
\varphi(\xi)=-\frac{1}{6N_2}\big( \sqrt{2} ( iN^{2/3}\sqrt{3}+12\xi N^{1/3}_1 a_1 + 
\end{equation}
$$
N^{2/3}_1 -4i\sqrt{3}-4N^{1/3}_1+4)^{1/2}
(-6\xi N^{1/3}_1 a_1   + N^{2/3}_1  +
$$
$$
+2 (N_1-4))^{1/2}
\big(i N^{2/3}_1 \sqrt{3}-12\xi N^{1/3}_1a_1
- N^{2/3}_1 -4i\sqrt{3}+ 4 N^{1/3}_1 -4
\big)^{1/2}\times F(\xi_0 | k),
$$
$$
\xi_0=(N^{2/3}_1-4)^{-1/2} \times
$$
$$
\times
\big(  (1/6-i/6) 3^{3/4}(i N^{2/3}_1 \sqrt{3} +12\xi N^{1/3}_1 a_1  + N^{2/3}_1 -4i\sqrt{3}
-4 N^{1/3}_1 + 4)^{1/2} \big),
$$
$$
k= \frac{(1+i) 3^{1/4} \sqrt{N^{2/3} - 4}}{\sqrt{i N^{2/3} \sqrt{3} + 3 N^{2/3} -4i\sqrt{3}+12 }}
$$
\noindent
where  $$N_1=12 \sqrt{3a_3}\sqrt{27 a_1^2 a_3 -4} a_1 -1-8 a_1^2 a_3 +8,$$
$$
N_2=\sqrt{i N_1^{2/3} \sqrt{3} +3 N_1^{2/3} -4i\sqrt{3} + 12 } N^{1/3} a_1 \sqrt{a_1 \xi^3 -\xi^2 +a_3}.
$$
 For a massless particle, the differentiation of formula (\ref{18}) (with  the aim of passing to a generalization of the Binet
 formula  \cite{Adler}) leads to the elimination of the term containing the constant $\Lambda$.
 This means that the curvature of a ray of light in
 the field of a massive point object (compact star, black hole) does not depend on it, and remains exactly the same as for the
 Hilbert--Schwarzschild  metric. However, for the case $m\neq 0$ in formulas (\ref{17}), the precession at pseudo--elliptical motion
 of the considered particle around the center with mass $M$ will be different; for this, its motion should be represented in
 terms of anomalous displacement $\zeta$, eccentricity  $\varepsilon$ and orbital parameter $\ell$ (see \cite{Adler}--\cite{Brumberg}):
  $\xi \ell=1+\varepsilon \cos\:\zeta$. Then, for the
  dependence $\zeta (\varphi)$, instead of the Abel equation, we   approximately obtain the differential equation for
  $d\zeta/d\varphi =\sum_{j=-2}^{j=1}q_j(E,L,\varepsilon,\ell) \zeta^j$.
  After its direct integration, we come to the value of the displacement of
   the aphelium for $K(=1,2,...)$ rotations:
 \begin{equation}
\delta \varphi_K= 2\pi K \bigg(1+\frac{3GM}{2c^2\ell}+\frac{3}{2}(1-\varepsilon^2)^{-1/2}
\big(
1-\frac{4GM}{\ell}+\frac{GM}{c^2 r_1}
\big)\big(    \frac{3GM}{\Lambda c^2 r_1^2 \ell }-1      \big)^{-1}
\bigg),
\end{equation}\noindent
where $r_1 (1\pm \varepsilon)$  are the
    semi--axes of the precessing orbit. From the last formula
    we see that the perturbation of the precession angle is linearly proportional to the value of the cosmological term.

\vskip 19pt

 {\bf{4.~Conclusions}}
\vskip 14pt

In this article we continue the studies  \cite{Fimin1}--\cite{Fimin3} on the analyzing of the influence of the
geometry of a gravitating body and of the possibility that the natural form of the  interacting mass objects law is a
combination of attraction and repulsion. Despite the fact that conclusions about this are made on the basis of
approximation of a weak gravitational field, generalization to strong fields and transition to the form of
Einstein's equations in general relativity seems to be confirmed by the observations. Moreover, this fact
naturally explains the existence of the cosmological term and its meaning. It is reasonable to proceed with
the analysis of particle motion in a modified gravitational field from the Hilbert metric, which leads to certain conclusions
regarding the motion test particle, and this can in principle be considered as the basis for the observer experiment.

In the next paper, based on the modified Newton's law discussed here,  we will consider the influence
of the law of gravity on the Friedman-type models, related with the validity of the homogeneous model of the expansion of the Universe.

This work was supported by the Russian Science Foundation, grant № 20-11-20165.

\end{document}